\documentclass{ws-ijicc}
\usepackage[comma]{natbib}
\usepackage{graphicx} 
\usepackage[hyphens]{url}
\usepackage[utf8]{inputenc}
\usepackage[english]{babel}
\usepackage{amsmath}
\usepackage{amsfonts}
\usepackage{amssymb}
\usepackage{booktabs}
\usepackage{comment}
\usepackage[amsmath,thref,thmmarks]{ntheorem}
\usepackage{listings}
\usepackage{tabularx}
\usepackage{float}
\usepackage[caption=false]{subfig}
\usepackage{upquote} 
\usepackage{eurosym} 
\usepackage[mathletters]{ucs} 
\usepackage{fancyvrb} 
\usepackage{tikz}
\usetikzlibrary{shapes.geometric,arrows}

\usepackage{xcolor}
\usepackage[xcolor]{changebar}

\definecolor{incolor}{rgb}{0.0, 0.0, 0.5}
\definecolor{outcolor}{rgb}{0.545, 0.0, 0.0}

\usepackage{soul}

\begin{document}


%
\catchline{}{}{}{}{}
%
%

\markboth{Bonart and Samokhina, et al.}{An Investigation of Biases in Web Search Engine Query Suggestions}

\title{An Investigation of Biases in Web Search Engine Query Suggestions}

\author{Malte Bonart}
\address{TH Köln - University of Applied Sciences, Cologne, Germany}

\author{Anastasiia Samokhina}
\address{Recogizer Group GmbH, Bonn, Germany}

\author{Gernot Heisenberg}
\address{TH Köln - University of Applied Sciences, Cologne, Germany}

\author{Philipp Schaer}
\address{TH Köln - University of Applied Sciences, Cologne, Germany}

\maketitle

\begin{abstract}

\textbf{Purpose} - Survey-based studies suggest that search engines are trusted more than social media or even traditional news, although cases of false information or defamation are known. In this study, we analyze query suggestion features of three search engines to see if these features introduce some bias into the query and search process that might compromise this trust. We test our approach on person-related search suggestions by querying the names of politicians from the German Bundestag before the German federal election of 2017. \\

\textbf{Design/methodology/approach} - This study introduces a framework to systematically examine and automatically analyze the varieties in different query suggestions for person names offered by major search engines. To test our framework, we collected data from the Google, Bing, and DuckDuckGo query suggestion APIs over a period of four months for 629 different names of German politicians. The suggestions were clustered and statistically analyzed with regards to different biases, like gender, party, or age and with regards to the stability of the suggestions over time.\\

\textbf{Findings} - By using our framework, we located three semantic clusters within our data set: Suggestions related to (1) politics and economics, (2) location information, and (3) personal and other miscellaneous topics. Among other effects, the results of the analysis show a small bias in the form that male politicians receive slightly fewer suggestions on "personal and misc" topics. The stability analysis of the suggested terms over time shows that some suggestions are prevalent most of the time, while other suggestions fluctuate more often.\\

\textbf{Originality/value} - This study proposes a novel framework to automatically identify biases in web search engine query suggestions for person-related searches. Applying this framework on a set of person-related query suggestions shows first insights into the influence search engines can have on the query process of users that seek out information on politicians.\\


\textbf{Paper type} - Research Paper\footnote{Preprint, accepted for publication, 27-Sep-2019, \textit{Online Information Review}, forthcoming, \textit{https://doi.org/10.1108/OIR-11-2018-0341}. This work is licensed under a Creative Commons Attribution-NonCommercial 4.0 International License. Any reuse is allowed in accordance with the terms outlined by the licence. For commercial purposes, permission should be sought by contacting permissions@emeraldinsight.com.}
\end{abstract}

\newpage

\section{Introduction}
\label{sec:introduction}
A recent survey by the PEW Research Center \citep{matsa_10_2016} showed that today most American adults still get their news from TV (57\%), but that online sources are a strong second with 38\% followed by radio (27\%) and print news (20\%). 
In younger cohorts online sources exceed TV news consumption and we can see a clear trend for a rising importance of online news sources.  
Politics are no exception to this change in information behavior. As paraphrased by Schiffmann: ``were it not for the internet, Barack Obama would not be president" \citep{schiffmann_reason_2008}. In light of the ongoing discussion on ``fake news'' and the impact of Donald Trump's online presidential election campaign \citep{parkinson_click_2016} the influence of the systems underlying online news and information sources should be discussed. 

Information found online, might influence real-world events such as the results of elections, especially if search engines manipulate the way information is queried. Past research studies have shown that biased search rankings can shift the voting preferences of undecided voters by up to 20\% and more, and such rankings can be masked in such a way that people show no awareness of the manipulation~\citep{epstein2015search}. While the presentation of the effect size is criticized for being exaggerated and falsely calculated \citep{zweig2017}, the study nevertheless shows the importance of looking for biases in online systems concerning their real-world impact. In pre-election phases shifting the voting preferences can have a particular influence on the future political scene of an entire country \citep{RePEc:cep:stieop:63}. The influence that search engines might have on events shows the importance of studying online search behavior.

Focusing on the influence of search engines is of high importance if we look at the level of trust people show in search engines. The 2017 Global Edelman Trust Barometer \citep{antoine_harary_2017_2017} asked for the trust in different sources of news and information. It ranked search engines as the most trusted information sources (64\%) followed among others by traditional media (57\%), and social media (41\%). 

\begin{figure}[t]
    \centering
    \includegraphics[width=.48\textwidth]{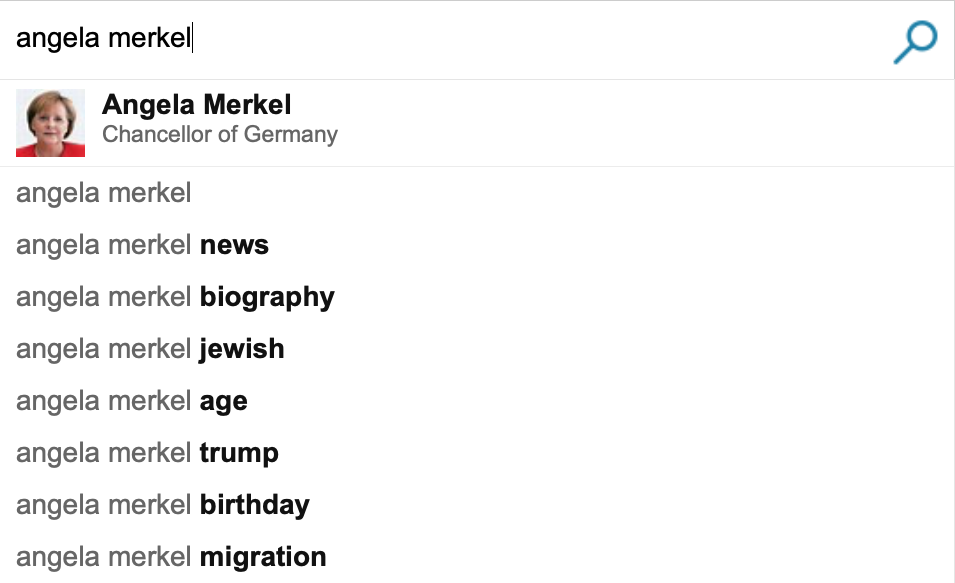}
    \includegraphics[width=.48\textwidth]{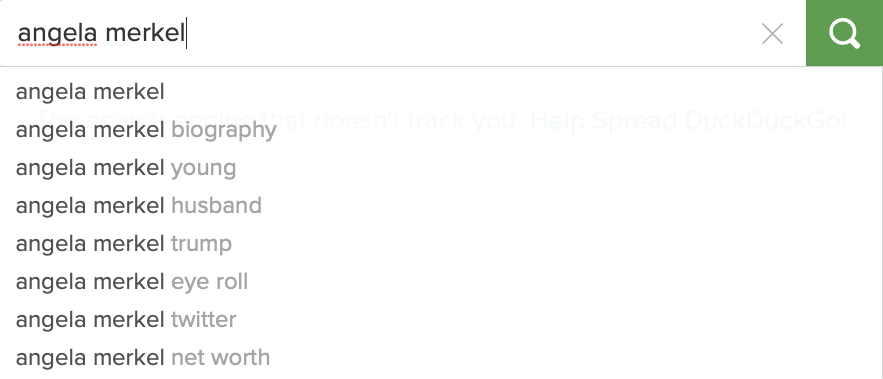}
    \caption{Two different query suggestions for the query on Angela Merkel (Bing on the left, DuckDuckGo on the right).}
    \label{fig:example}
\end{figure}

Specifically for person-related searches, we ask if this trust is compromised by biases inherent to the underlying retrieval, ranking, and suggestion algorithms. In the following paper, we present a general query suggestion data acquisition, processing, and analysis framework to automatically identify biases in an exploratory manner. An example of different query suggestions is shown in Figure~\ref{fig:example}.  

We concentrate on the following research questions:

\begin{description}
    \item[RQ1] How can we automatically identify clusters and patterns in web query suggestions for person-related searches?
    \item[RQ2] To what extent can metadata on the persons searched (e.g., gender, age, party membership) be used to explain possible biases?
    \item[RQ3] How can we measure and analyze the stability and persistence of possible biases in the suggestions? 
\end{description}

We approached these research questions by focusing on the tasks of identifying any systematic patterns in the query suggestions (RQ1) and by concluding how these suggestions are generated (RQ2). As the crawled data sets typically contain query suggestions from a longer period, we were also interested in the long-term fluctuation of the suggestion terms (RQ3).

We intend to contribute to the topic of information biases in search engines by applying a text mining approach. We focus on the statistical identification and quantification of systematic patterns in the data. Whether possible biases do influence the process of political opinion formation of individuals, is not part of this work.

In an experimental study, we applied our framework to the case of online searches for German politicians: We gathered and analyzed query suggestion data for the names of politicians from three popular web search engines during four months before the German federal election 2017. We used the names of the politicians of the German Bundestag (the German federal parliament) and enriched the crawled data with socio-demographic information, such as age, gender, home state, and party membership.

We contribute to the research by presenting a novel framework to automatically identify biases in web search query suggestions for person-related searches. We do not study query suggestions for individual cases but systematically for many search terms and over a long period. The presented methods do not need access to the search engine's query log or its underlying algorithms. Therefore, our approach contributes to the research on algorithm audits for search engines \citep{sandvig2014auditing}. In contrast to other bias quantification approaches, our methodology is data-driven and does not require a normative reference scheme against which a bias is measured (e.g., by defining concepts of sexism, political leaning or offensive speech).

The paper is structured as follows: In Section~\ref{sec:literature}, we give a literature overview on query suggestions, the topics of biases in search engines and the impact on the search and decision processes of users. In Section~\ref{sec:methods}, we describe the general framework for the identification of biases. It consists of data acquisition, preprocessing, and analysis modules. In the following Section~\ref{sec:casestudy}, we apply the framework on searches for German politicians before the federal election in 2017. In Section~\ref{sec:discussion}, we discuss the results of the bias and stability analysis and conclude in Section~\ref{sec:conclusion}.

\section{Literature Study}
\label{sec:literature}
Search engines support their users by suggesting possible completions of their partially typed queries. Typing longer queries with more keystrokes increases the possibilities for inaccurate query formulations or typos. Therefore, the goal of the suggestions is first, to speed up users' searches by reducing the number of typing interactions and second, to avoid the wording problem. Usually, these suggestions are shown as drop-down menus or are inline suggestions next to the top results of the search engine result page (SERP). Different types of suggestions are possible, ranging from a single word or \emph{term suggestion} to full \emph{query suggestions} or \emph{query auto completions} \citep{chen_term_2011, cai_survey_2016}.  

In today's web-based systems, algorithms generate suggestions by analyzing the query log files to find the most popular queries. Although query suggestions are available on all major platforms, details on how search engines implement the suggestions are not available. Google describes their query ``prediction'' method as being based on the frequency with which other users have searched for the suggested term, the location and language of the user and the popularity of recent trending searches ~\citep{google-autocomplete,mysen_dynamic_2011}. 

Different studies based on the AOL logs showed the positive effects of query suggestions on the search process like fewer empty result sets, fewer query drifts due to a more precise query wording of using the terms of the corpus, and a generally quicker search process \citep{pass_picture_2006}. \citet{kato_when_2013} showed that users rely on query suggestions ``(1) when the original query is rare, (2) when the original query is a single-term query, (3) when query suggestions are unambiguous, (4) when query suggestions are generalizations or error corrections of the original query, and (5) after the user has clicked on several URLs in the first search result page''. 

Query suggestions are specially designed to support users that are uncertain in their information need, and that find themselves in a so-called Anomalous State of Knowledge (ASK) \citep{belkin_ask_1982}. In this state, they are uncertain of their own genuine information need or the right way to formulate and verbalize it. In retrieval settings where users are less familiar with the given search topic, they tend to consult external knowledge representation systems \citep{DBLP:conf/ercimdl/HienertSSM11}. In addition to this, \citet{kelly_comparison_2009} points out that query suggestions seem particularly important in cases where users are searching for topics on which they have little knowledge or familiarity. As query suggestions are meant to support the uncertain user, the question arises if a biased suggestion mechanism can spoil this uncertainty. 

\citet{friedman1996} were among the first who studied the phenomenon of biases in computer systems in general. They defined bias as systematic and unfair discrimination against (groups of) individuals. The process is unfair if those individuals receive a negative and inappropriate outcome. It is systematic if it does not occur randomly. For the measurement of biases in online retrieval systems, bias is defined as a systematic, inclusion, exclusion, or prominence in the selection or the content of the retrieved documents \citep{introna2000, mowshowitz2002}. In a narrower statistical sense, errors in the estimation or sampling process cause a systematic deviation from the true unknown distribution \citep{baeza-yates_bias_2018}. It is important to note that biases can only be relatively measured as a systematic deviation from a reference that is a fair norm or ideal distribution. 

The many forms of biases that exist in the literature are due to the different underlying references utilized. \citet{mowshowitz2005} approximates the true distribution by combining the output of several comparable retrieval systems (e.g., a set of other web search engines). \emph{Coverage-bias} measures the share of web pages that are indexed by a particular search engine compared to an independent web crawl \citep{vaughan2004}. \citet{fortunato2006} study the \emph{popularity-bias}, the assumption that search engines amplify the popularity of all-ready well-known pages. Here, a website's traffic is compared to a theoretically deduced traffic that would occur without the impact of search engines. \emph{Topical bias} is concerned with the actual content of retrieved web documents \citep{pitoura_measuring_2018}. \citet{kulshrestha_search_2018} analyzes the political leaning of a ranked retrieval set against an independently crawled set that represents the ground truth. \emph{Political bias} of search engines was also measured by linking the URLs from a ranked list of documents to social media utterances of Twitter users whose past voting decision was known \citep{robertson_auditing_2018}.

These studies focus on biases in search engine result pages, and less research exists on slanted query suggestions. \citet{noble_algorithms_2018} collects and reports many cases of sexism and racism in search suggestions. \cite{yenala_convolutional_2017} uses supervised machine learning methods to detect inappropriate suggested queries. The success of the method relies on the quality and quantity of manually labeled training examples. As \citet{hiemstra_query_2017} points out, there are adverse effects that complement the general positive image of query suggestions. 

Our framework, presented in the next section, detects systematic topical biases in query suggestions related to a set of search terms with common attributes. The topics are derived from the data itself in an unsupervised manner. Systematic biases are then measured as deviations from the distribution of topics. On an abstract level, this approach is comparable to the bias studies in \citet{mowshowitz2005}. If groups of search terms with similar characteristics deviate from the overall topical distribution, this group is reported as having biased search suggestions. As we do not impose any prior concept of fairness or ideal distribution, our methods should only be applied in an exploratory manner. It is the researcher's responsibility to interpret the topics derived from the data and the reported biases afterward.

Manipulated and biased rankings can have effects on the user. \citet{otterbacher_investigating_2018} look at gender biases in image search and how the users' perception can reinforce stereotypes. In their well-known experiments, \citet{epstein_suppressing_2017} present manipulated rankings of political result pages and measure changes in the voting preferences of the users. The effect of manipulated search engines was also confirmed for medical-related searches in the context of vaccination beliefs  \citep{allam_impact_2014}. In this paper, we neglect the users' perspective and focus on the task of bias detection and measurement.

\section{Automatic Identification of Biases in Web Query Suggestions}
\label{sec:methods}
\begin{figure}[t]
    \centering
    \includegraphics[width=\linewidth]{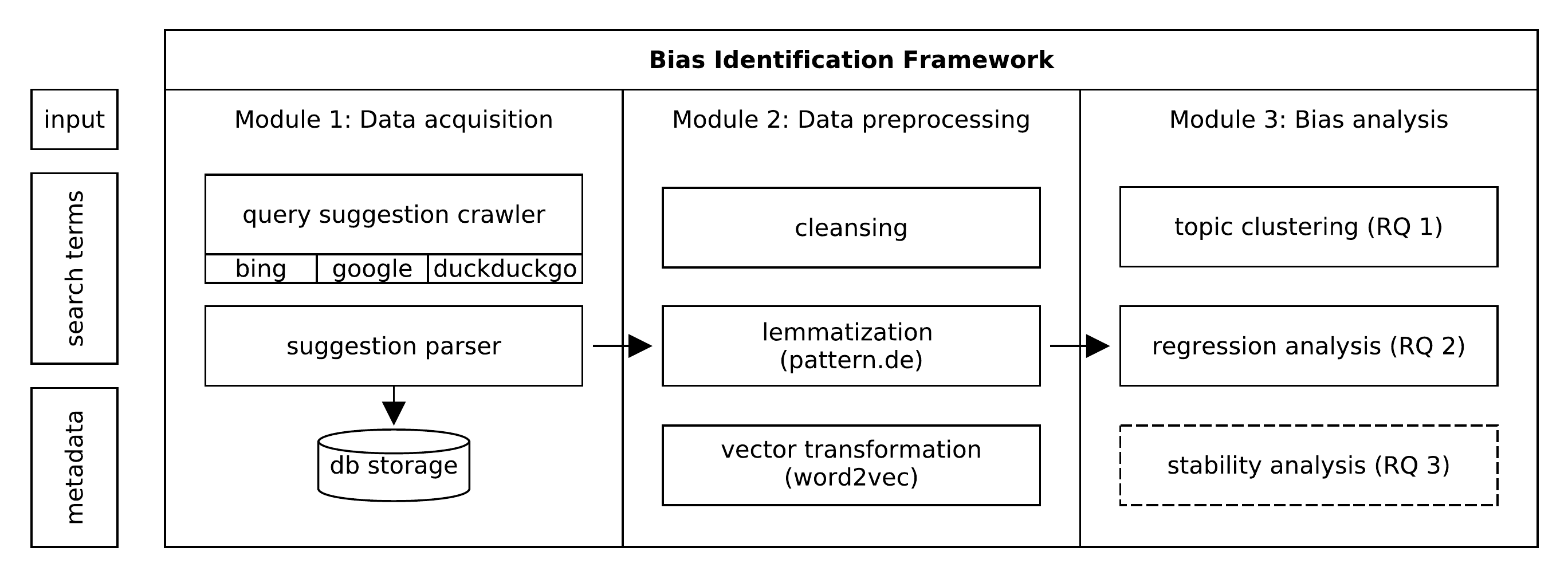}
    \caption{Bias identification framework with the three modules for acquiring, processing and analyzing query suggestions from different web search engines. The analytic procedures in the last module address the research questions RQ1-RQ3 as proposed in section \ref{sec:introduction}.}
    \centering
    \label{fig:pipeline}
\end{figure}

This section describes the framework to detect and analyze potential biases in query suggestions (see Figure~\ref{fig:pipeline}). It consists of three modules that can be applied to different use cases. In this prototype, we relied on text mining libraries for German language corpora, but the different procedures can be easily modified to address other languages as well. In the following, we will describe each of the three modules.

\subsection{Module 1: data acquisition}

The bias analysis is based on a set of initial search terms $t_i$ with $i = 1, \dots, N$ which share a set of $P$ meta-attributes $\{x_{i,1}, \dots, x_{i,P}\}$. For example, $t_i$ can be the name of a prominent personality with some meta-attributes such as \emph{age}, \emph{gender} or \emph{country of birth}. A specialized web crawler was developed to fetch the query suggestions related to the search terms over a longer period. It is capable of systematically browsing the auto-complete API of three different search engines: Google, Bing, and DuckDuckGo. This is done by sending an HTTP request with the search term to each search engine twice per day. 

The HTTP response of each search engine API delivers a list of query suggestions with a different number of items typically ranging from 4 to 10. Both the original query term and the corresponding query suggestions are stored in a database. A script parses the raw request data and extracts the suggestions for further processing. Most of the suggestions are of the form \texttt{\{<search term>+<suggestion>\}}. Here, the search term entered into the input field is similar to the prefix of the suggestion. In this case of true auto-completion, the search term was stripped off. An example of the parsed suggestions in the database for the term ``Angela Merkel'' and from the English DuckDuckGo API is shown in Table~\ref{tab:dataset}. 

\begin{table}[t]
\caption{Illustration of the query suggestions data set retrieved by the DuckDuckGo API.}
\centering
\begin{tabular}{c@\quad c@\quad c@\quad c@\quad c}
  \toprule
 session id & queryterm & datetime & suggestterm & position \\ 
  \midrule
2086154 & Angela Merkel & 2017-08-03 05:46:17 & biography &   0 \\ 
2086154 & Angela Merkel & 2017-08-03 05:46:17 & husband &   1 \\ 
2086154 & Angela Merkel & 2017-08-03 05:46:17 & young &   2 \\ 
2086154 & Angela Merkel & 2017-08-03 05:46:17 & trump &   3 \\ 
2086154 & Angela Merkel & 2017-08-03 05:46:17 & meme &   4 \\ 
2086154 & Angela Merkel & 2017-08-03 05:46:17 & eye roll &   5 \\ 
2086154 & Angela Merkel & 2017-08-03 05:46:17 & twitter &   6 \\ 
2086154 & Angela Merkel & 2017-08-03 05:46:17 & photo &   7 \\ 
2086154 & Angela Merkel & 2017-08-03 05:46:17 & snl &   8 \\ 
2086154 & Angela Merkel & 2017-08-03 05:46:17 & and donald trump &   9 \\
  $\cdots$ & $\cdots$ & $\cdots$ & $\cdots$ & $\cdots$ \\
\bottomrule
\end{tabular}
\label{tab:dataset}
\end{table}

\subsection{Module 2: data preprocessing}

The second module of the framework deals with cleaning the suggestions, with the problem of mapping single query suggestion terms to a base word form and with the construction of a high dimensional word vector that represents the semantic meaning of the query term. This allows for the grouping of stand-alone words into semantic clusters.

The suggestions are cleaned from special characters, numbers, and German umlauts. In order to tackle the word form problem, we rely on lemmatization. The number of existing German implementations for lemmatization is limited, and we used the text analysis module \emph{pattern.de} \citep{gesmundo2012lemmatisation} that contains a built-in parsing function. It annotates each word with its base form. The language model the parser is built on is described in \citet{schneider1998adding}.

The process of vector transformation implies taking a corpus of text as input and building a vector space of several hundred dimensions. Each unique word in the input is assigned a corresponding vector in that space. \citet{rehurek_lrec} provide a Python re-implementation of the \emph{word2vec} model~\citep{DBLP:journals/corr/abs-1301-3781} that is used in this module. We relied on the ``GermanWordEmbeddings'' toolkit as a pre-trained model. Its corpus included more than 600,000,000 words and was trained with the German Wikipedia and news articles written in German~\citep{germanwords}. 

\subsection{Module 3: data analysis}

\subsubsection{Cluster analysis}
After transforming each query suggestion into a high dimensional numeric word vector, an unsupervised cluster analysis was applied. For words with a similar meaning, the distance between their word vectors is small. Therefore, by clustering the vectors, semantically similar words are grouped. Clustering enables the identification of topical categories in the search terms' suggestions. The k-means clustering algorithm was chosen here \citep{Kanungo02anefficient}. The number of clusters $K$ is determined by relying on established heuristics such as the total within-cluster variation~\citep{Pelleg00x-means:extending}, the Silhouette or the Calinski-Harabaz score~\citep{Rousseeuw:1987:SGA:38768.38772}. 

For each initial search term $t_i$ with $n_i$ unique query suggestions, we can sum up the relative number of its suggestions belonging to each cluster topic. This provides a measure of how well each topic is represented in the suggestions for this search term. Hence, consider the cluster score $y_{i, k}$ with $k \in \{1, \dots, K\}$. It is defined as $y_{i,k} = \frac{1}{n_i} \sum_{j = 1}^{n_i} \bm{1}(s_{i,j} = k)$, where $s_{i,j}$ is the topical category of the $j^{th}$ suggestion and the $i^{th}$ search term. Since the cluster score is larger for search terms with more query suggestions, we normalized the score by the term's total number of query suggestions $n_i$. This results in a relative cluster score in the range $y_{i,k} \in [0,1]$.

\subsubsection{Regression analysis}
\label{sec:regression}

The proportional cluster scores represent the distribution of a search term's suggestions over the topical categories. Significant differences in the cluster distributions for a group of search terms can indicate possible systematic biases. For example, in the case of searches for the names of prominent personalities, the cluster analysis could reveal two clusters: A cluster with suggestions related to family members (e.g. \emph{mother, brother, daughter, ...}) and a cluster related to social media and other online resources (e.g. \emph{wiki, twitter, instagram, ...}). Suppose that compared to males, female personalities had a significantly higher score for the first cluster, meaning that more suggestions from the first topic show up if users search for them online. Our framework is capable of discovering these kinds of systematic biases.

To detect significant deviations in the cluster scores regarding the meta-attributes of the search terms, a multiple regression analysis was performed. The cluster scores for one cluster $y_{i, k}$ over all search terms form the independent variable and the meta-attributes $x_{i,p}$ are the input to the model. Since we were only interested in detecting broad systematic patterns we  relied on the standard linear regression model for the ease of interpretation \citep{hastie_elements_2001}. Formally, for some clusters $k \in {1, \dots, K}$ the model can be represented by the regression equation (Equation \ref{eq:regression}):

\begin{equation}
y_{i, k} = \beta_0 + \beta_1x_{i, 1} + \dots + \beta_Px_{i, P} + \epsilon_i,
\label{eq:regression}
\end{equation}

where $\epsilon_i$ is the independent and identically distributed error term and $i = 1, \dots, N$ are the observations indices. A significant positive or negative coefficient $\beta_p$ then indicates a biased topical distribution for this cluster and the attribute $x_{i, p}$. 

\subsubsection{Stability analysis}
\label{sec:stability}

The cluster scores in the regression analysis are calculated by considering the unique query suggestions over the entire observation period. By relying on such a cross-sectional analysis, every query suggestion is assigned an equal weight. A suggestion that appears only once during the observation period is assumed to show the same importance as a suggestion which is prevalent most of the time. To see whether the time independence assumption is reasonable, it is important to check for the degree of stability of the rankings. 

As a supplement to the bias identification framework, we separately analyze the long-term ranking stability of the query suggestions gathered in module 1. This is done by comparing each term's suggestion ranking at each time-point with its earliest ranking in the sample and its respective preceding ranking. Formally, rankings of query suggestion lists are incomplete, top-weighted, and indefinite. The importance of the items decreases with the depth of the list. To compare the ranking of two lists, we rely on the rank-biased overlap ($RBO$) statistics \citep{webber2010}. It is capable of measuring the similarity of two indefinite and possibly non-conjoint rankings. The statistics can take values between $0$, indicating no similarity, and $1$ for identical lists. The degree of top-weightiness, hence the importance of items at the start of the list, can be controlled by the parameter $p$. For $p \rightarrow 0$ the first item in both lists gets assigned all the weight while for $p \rightarrow 1$ all list items become equally important.

\section{The Search for Politicians -- The Case of the German Bundestag in 2017}
\label{sec:casestudy}
In the previous section, we described the framework for exploratory bias detection on an abstract level. Here, we apply it to the names of German politicians and their respective suggestions in web search engines. The underlying data is archived in Zenodo \citep{dataset2019}.

Political activity on the web is already generating massive amounts of data, such as individuals’ political conversations, donations, online formats of news and politics, political blogs, online public speeches and openly available information about political activities and involved individuals. Hence, the search for politicians is a common task as citizens might inform themselves about a candidate before elections. However, as most persons with publicity, politicians do not always become the object of public interest only because of their primary work, but due to other aspects such as their private life, family affairs or fashion style and appearance. 

\subsection{Data acquisition and processing}

First, the data acquisition phase was initiated by collecting the names of all politicians of the $18^{\text{th}}$ German federal parliament (German Bundestag) in the constitution of late 2016. Additional socio-demographic information for each politician, such as the age, the home town, and the party affiliation, was collected too. This information corresponds to the framework's meta-attributes, for which a possible bias can be detected. The data set included $36\%$ female politicians. The average age was $54$. Most members of the parliament originated from the German federal state of North Rhine-Westphalia ($22\%$), which is the largest populated, while the smallest amount of members of the parliament originated from the German federal state of Bremen ($1\%$). The biggest party in the sample was the CDU (``Christian Democratic Union") with a proportion of $40\%$. Since there was only one member in the group ``non-attached member" in the \textit{party} category, the politician of this group was excluded from further analysis.

\begin{figure}[t]
	\centering
	\includegraphics[width = 0.5\textwidth]{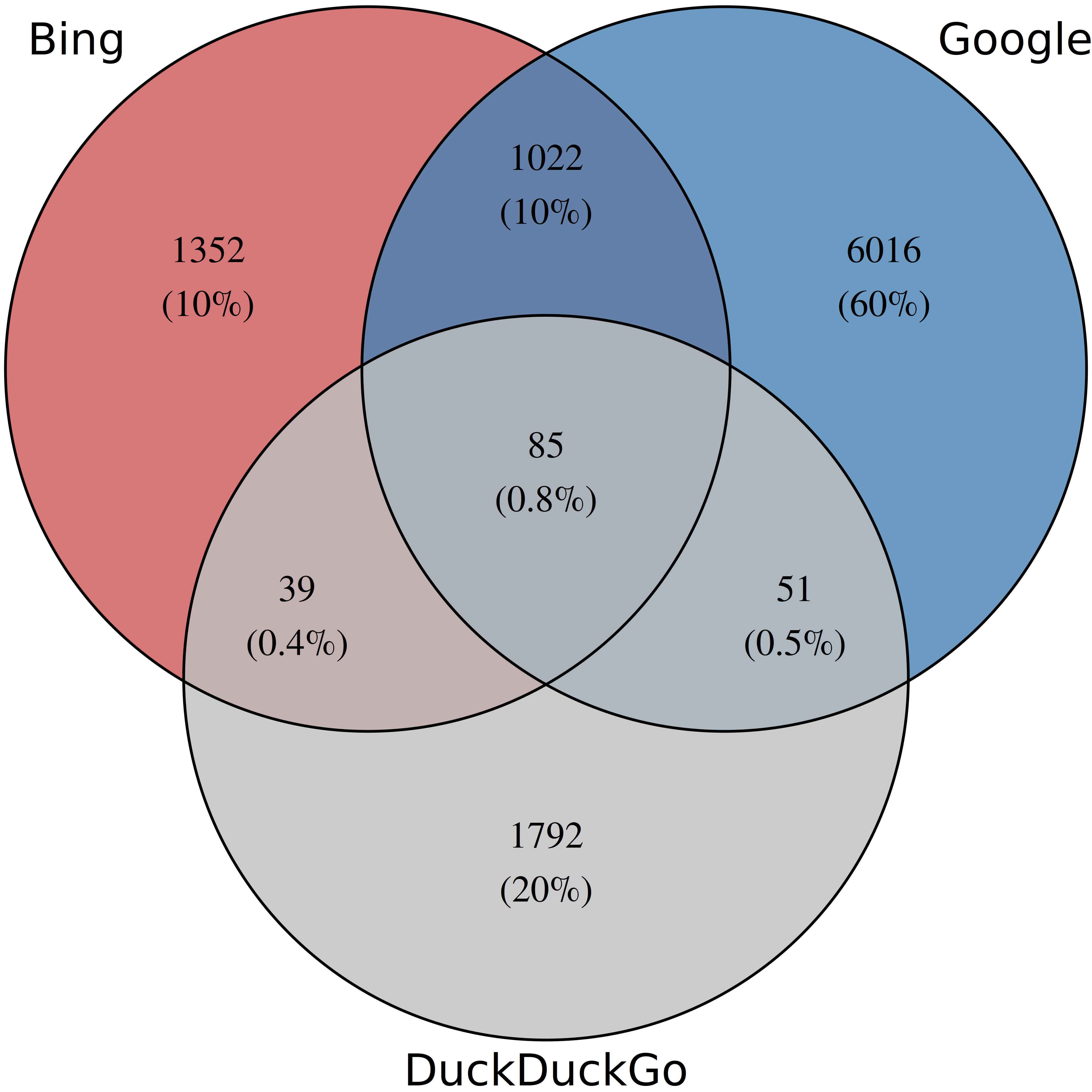}
	\caption{Venn diagram with the sizes of the sets of overlapping and unique search terms for each of the three search engines. The total number of unique search terms is $10,357$.}
	\label{fig:engineoverlap}
\end{figure}

Searches for Bing and Google were carried out using the German language setting. English settings have been used in the DuckDuckGo search engine. The data was crawled for four months between $2017/02/13$ and $2017/06/19$. The data set of query suggestions included more than $10,000$ unique suggestions from all three search engines. Regarding the query suggestions for the politicians, we observed some differences between the search engines (see Figure~\ref{fig:engineoverlap}): First, the number of unique search terms was largest for Google, indicating a higher diversity in terms for this provider. Secondly, the three search engines had all small overlapping sets of suggestions with only $85$ unique suggestion terms shared by all three engines. Therefore, by combining the three different search engines and their query suggestions, we could create a larger and diverse data set with more unique suggestions.

After cleaning and lemmatization, $3,824$ unique single word suggestions were leftover, and the vector-transformation algorithm could identify $3,040$ words. Word embedding vectors for each of the suggestions were used to perform a cluster analysis. The heuristics suggested a cluster size of $K = 3$. By manually evaluating the clusters, we assigned a label that best describes the topic of each cluster (see Table~\ref{tab:my_label}). 

The cluster analysis provided useful insights into the topical distribution of the search suggestions. The first cluster contained suggestions referring to political and economic topics. These suggestions were related to the formal position and to trending political news events during that time. The second cluster included many suggestions related to personal and emotional topics, including references to tabloid journalism. The final cluster solely consisted of names of cities and regions. We assume, they possibly refer to the politicians' electoral districts and the appearances in election campaigns.

\begin{table}[t]
    \caption{Selected query suggestions from each of the three clusters identified by the k-means procedure. The terms were translated from German to English.}
    \setlength{\tabcolsep}{8pt}
    \begin{tabular}{p{0.29\textwidth}p{0.29\textwidth}p{0.29\textwidth}}
    \toprule
    \textbf{cluster 1} & \textbf{cluster 2} & \textbf{cluster 3} \\
    $717$ suggestions related to politics and economics & $1,251$ suggestions related to personal and emotional topics & $1,072$ suggestions containing location information (cities) \\
    \midrule
    parliamentary office & pregnant & giessen  \\
    prosecutor & toupet & kassel  \\
    minister of transport & cartoon & radeberg \\
    integration commissioner & simple & rudolstadt \\
    member of bundestag & wife & heiligenhafen\\
    minister & brother & forchtenberg \\
    treasurer & tie  & fulda\\
    middle east policy & airplane & moers \\
    $\cdots$ & $\cdots$ & $\cdots$ \\
    \bottomrule
    \end{tabular}
    \label{tab:my_label}
\end{table}

\subsection{Bias identification}

The cluster assignments were used to calculate the topical distribution for each politician. We then performed a regression model on the proportion of unique suggestions belonging to cluster 1 and cluster 2. Cluster 3 containing German geographical locations did not promise any insights, as users searching for geographical locations in combination with the names of politicians could have been querying for the place of birth, the electoral district or other topics that are out of reach without the specific context. Therefore, we focussed further investigations on clusters 1 and 2 only. For groups with similar socio-demographic characteristics, we tested whether the group status had a significant positive or negative effect on the respective topical distribution. In this case, our framework reports a bias for this group (compare to section \ref{sec:regression}). 

\begin{table}[tp]
\caption{Linear Regression model for fitting the proportion of the politicians' query suggestions belonging to cluster 1 or cluster 2. Shown are the coefficients along with the significance value of the test for non-zero coefficients, the F-test for overall significance and the adjusted $R^2$ measure. Statistically significant values ($p < 0.05$) are marked bold.}
\centering
\begin{tabular}{rcccc}
  \toprule
  & \multicolumn{4}{c}{Dependent variable}\\[1em]
   & \multicolumn{2}{p{3cm}}{\centering Cluster 1: Politics and Economics} & 
   \multicolumn{2}{p{3cm}}{\centering Cluster 2: Personal and Emotional} \\[1.5em]
 Independent variable~ & Estimate~ & P-value \hspace{2em} & Estimate~ & P-value \\ 
 \cmidrule(r){1-1}  \cmidrule(lr){2-3} \cmidrule(l){4-5} 
$\beta_0$  & \textbf{0.242} & 0.000 & \textbf{0.353} & 0.000 \\ 
  Gender Male & -0.009 & 0.371 & \textbf{-0.024} & 0.025 \\ 
    Age & 0.000 & 0.396 & \textbf{-0.002} & 0.000 \\ 
   \cmidrule(r){1-1}  \cmidrule(lr){2-3} \cmidrule(l){4-5} 
State Baden-Württemberg & \multicolumn{4}{c}{reference category} \\
  State Bayern & 0.008 & 0.720 & -0.033 & 0.198 \\ 
  State Berlin & 0.001 & 0.968 & 0.032 & 0.244 \\ 
  State Brandenburg & -0.025 & 0.385 & -0.060 & 0.064 \\ 
  State Bremen & -0.033 & 0.483 & \textbf{-0.111} & 0.035 \\ 
  State Hamburg & -0.004 & 0.904 & 0.011 & 0.767 \\ 
  State Hessen & -0.022 & 0.303 & -0.034 & 0.145 \\ 
  State Mecklenburg-Vorpommern & \textbf{0.079} & 0.023 & -0.071 & 0.065 \\ 
  State Niedersachsen & -0.001 & 0.968 & -0.002 & 0.914 \\ 
  State Nordrhein-Westfalen & 0.000 & 0.997 & 0.005 & 0.770 \\ 
  State Rheinland-Pfalz & -0.004 & 0.874 & -0.001 & 0.967 \\ 
  State Saarland & -0.047 & 0.226 & -0.060 & 0.169 \\ 
  State Sachsen & -0.024 & 0.297 & -0.016 & 0.526 \\ 
  State Sachsen-Anhalt & -0.002 & 0.946 & -0.038 & 0.231 \\ 
  State Schleswig-Holstein & 0.028 & 0.281 & -0.026 & 0.373 \\ 
  State Thüringen & -0.008 & 0.792 & -0.019 & 0.556 \\ 
    \cmidrule(r){1-1}  \cmidrule(lr){2-3} \cmidrule(l){4-5} 
 Party CDU & \multicolumn{4}{c}{reference category} \\
  Party CSU & -0.019 & 0.463 & \textbf{0.057} & 0.048 \\ 
  Party DIE LINKE & 0.010 & 0.531 & -0.009 & 0.635 \\ 
  Party GRÜNE & 0.015 & 0.350 & \textbf{0.039} & 0.033 \\ 
  Party SPD & -0.006 & 0.598 & -0.004 & 0.734 \\ 
 \cmidrule(r){1-1}  \cmidrule(lr){2-3} \cmidrule(l){4-5} 
F-statistic & 0.865 & 0.638 & \textbf{2.944} & 0.000 \\
Adjusted $R^2$ & \multicolumn{2}{c}{0.000} & \multicolumn{2}{c}{0.061} \\ 
\midrule
Degrees of freedom & \multicolumn{4}{c}{605} \\
   \bottomrule
\end{tabular}
\label{tab:regression}
\end{table}

Table \ref{tab:regression} shows the regression results for cluster 2, which consists of query suggestions related to personal and emotional topics, and cluster 1, which comprises suggestions related to politics and economics. We found that the model for cluster 1 was not statistically significant. Since its $p$-value was quite large, the F-test could not reject the joint null hypothesis that all coefficients are zero. 

Hence, we focused on the significant second model, which predicts the proportion of query suggestions belonging to the cluster with terms related to personal and emotional topics. The base category for the attribute \emph{constituency} is ``Baden-Württemberg". The base category for the attribute \emph{party} is ``CDU". As an illustration, consider a $ 40$-year-old female politician from ``Baden-Württemberg" belonging to the party ``CDU": This model predicts an average cluster value of
$0.353 + 40\times (-0.002) = 0.273$. Hence, approximately $27\%$ of the unique query suggestions for this hypothetical politician belong to the cluster associated with personal and emotional terms. 

We found that both the gender and age, are negatively associated with the amount of personal and emotional query suggestions. Compared to an average female politician of the Bundestag, a male politician had on average $2.4$ percentage points fewer suggestions in cluster 2. Neglecting the fact that some suggestions were more visible than others and assuming that the performance of the word-embedding method works equally well for all types of words, this indicates that users of search engines were slightly more exposed to topics related to personal and emotional terms in combination with a female politician. Nevertheless, the effect is rather small.

Roughly, the same negative effect can be seen for the age of a politician: Keeping the gender or other socio-demographic variables constant, on average, the number of queries belonging to cluster 2 decreases by $2$ percentage points by steps of $10$ years. The most substantial effect was found regarding the constituency of the politician: A politician from the federal state of Bremen has 11 percentage points fewer suggestions in cluster 2 compared to the base category. However, as this state is the smallest in Germany, only $6$ politicians were included in our sample. Finally, we also found a positive effect for the parties ``Die Grünen" (The Green Party) and ``CSU" (The Christian Social Union in Bavaria). Compared to a politician from the ``CDU'', on average, a politician from ``Die Grünen" had roughly $4$ percentage points more suggestions in cluster 2.

While our proposed framework showed a possible way of identifying biases, we could only find weak effects in this specific use case. The overall level of variance was significant: The R-squared statistics showed that only around $6\%$ of the variation in the dependent variable was explicable by the independent model variables for the second cluster. While there were some interesting significant effects, e.g., for the gender of the politician, the effect size was rather small. The model for the first cluster showed no overall significance at all. 

\subsection{Stability of rankings over time}
\label{sec:case-stability}

In the last analysis step, we tested the long-term stability of the query suggestions in Google for the German politicians over time as mentioned and introduced in section \ref{sec:stability}. For each measurement date, we calculated the rank-biased overlap similarity of each ranking with its first and its previous ranking in the sample. The former gives insights into the overall stability, whereas the latter approximates a rate of change of the suggestions. The median of the suggestions for the lists was $9$ with a maximum of $10$ list entries. Therefore, for the calculation of the RBO score, we chose the weighting parameter to be $p = 0.90$. Doing so puts most of the weight ($83\%$) to the first nine list items and results in an expected evaluation depth of $10$ items as shown by \citet{webber2010}.

\begin{figure}[t]
       \includegraphics[width=\textwidth]{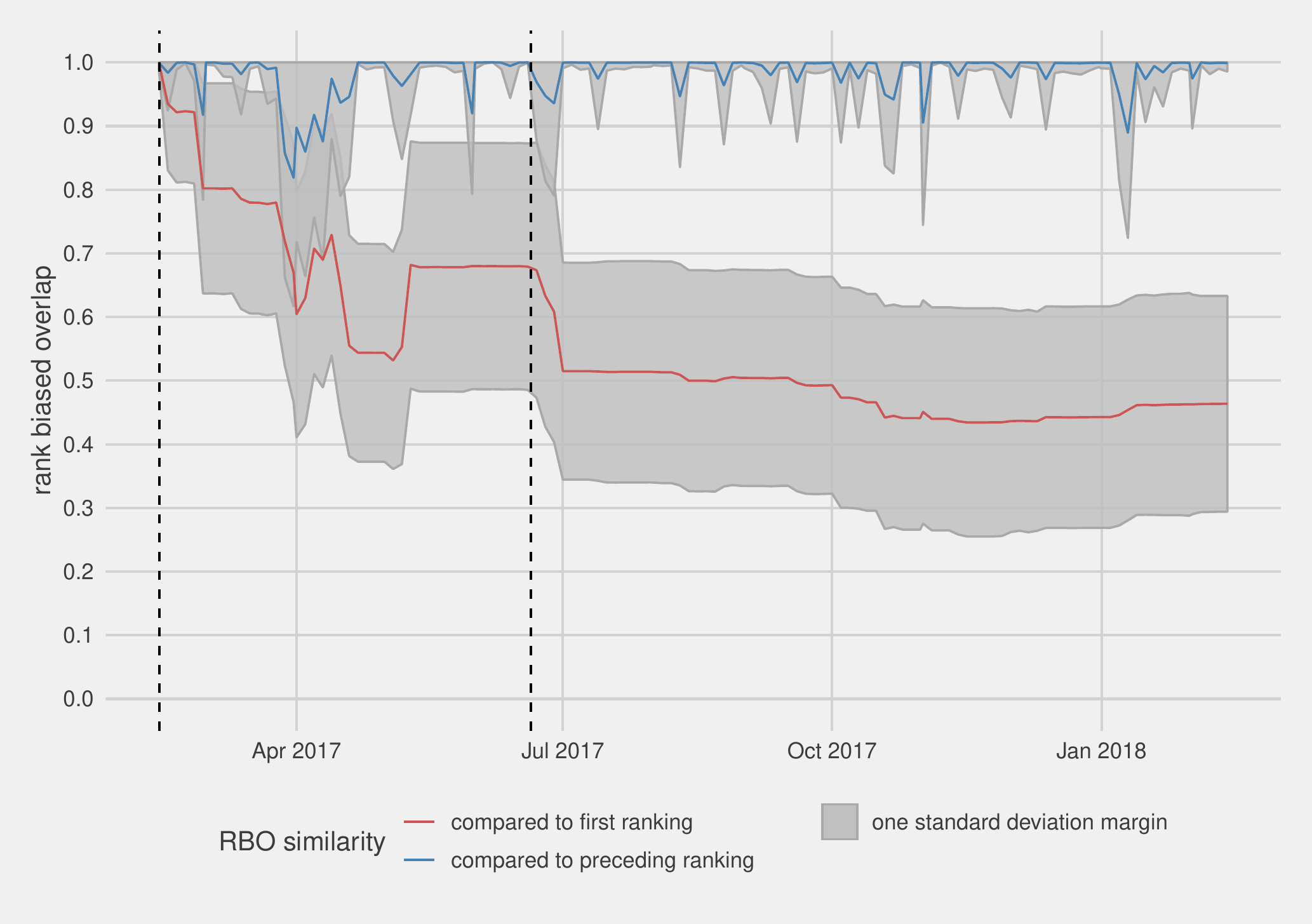}
     \caption{The colored lines represent the three day average for the politician's stability statistics (RBO value). The dashed lines indicate the crawling period for the suggestions used in the section on bias identification. The shaded areas are a standard deviation margin around the mean.}
     \label{fig:stability2}
\end{figure}

Figure~\ref{fig:stability2} shows the aggregated three-day average over the rankings of all politicians. The shaded area indicates the variability of the data in the form of a standard deviation margin around the mean. The overall similarity decreased over time. Nevertheless, after one year, the average similarity roughly converged to $\overline{RBO}=0.45$.  In early $2018$, each ranking compared to the ranking at the beginning still had around $45\%$ of their suggestions in common. As the figure shows, the rankings seemed to change faster at the beginning of the observation period. After July 2017, the RBO decreased only slowly from around $0.5$ to $0.4$. This finding indicates that some query suggestions were much more prevalent, while other suggestions fluctuated and exchanged more often. The successive rate of change was small, having a stability value of $\overline{RBO}=0.98$ on average. This rate stayed relatively constant, despite some negative shocks at several time points. Possibly, these temporal shocks occurred due to political events or changes in the Google query suggestion algorithms.

\section{Discussion}
\label{sec:discussion}
This work is the first that systematically investigated biases in web query suggestions for person-related searches. We proposed an analytic framework to identify possible biases in a data-driven and exploratory way and applied it to the case of web searches for German politicians. The modularity of the presented framework simplifies possible extensions and improvements of specific procedures. Based on the research questions formulated in the introduction, in this section, we will discuss the outcomes and limitations of our work and suggest starting points for further research. 

\subsection{Topical clustering of query suggestions}

\emph{RQ 1: How can we automatically identify clusters and patterns in web query suggestions for person-related searches?}

The query suggestions are grouped into topical clusters by first transforming single words into a high dimensional word vector that represents the semantic meaning of the word. Second, a clustering algorithm classifies geometrically close vectors, and therefore semantically similar words together. Future research can work on the improvement of the utilized word embedding methodology. This includes the implementation and evaluation of other embedding models such as \textit{fastText}, which can also process unseen and misspelled words not included in the training data \citep{bojanowski2016enriching}. In the current framework, only single word suggestions are clustered. By averaging or summing up the corresponding word vectors for longer query suggestions, it is possible to derive a vector representation for complete queries \citep{zamani2016}.

In the experiment on searches for the names of German politicians, the cluster analysis identified three distinct groups of semantically similar query suggestions. The three clusters were related to formal topics from politics and economics to more personal and emotional issues, and the names of cities and regions in Germany. Thus, regarding the first research question, for this case, our methodology has been proven successful. Further experiments are needed to evaluate the general validity and applicability of the clustering methodology. Promising use cases might be the search for prominent personalities, well-known business persons, and politicians participating in other national elections. For the clustering itself, we determined the number of topics by relying on several data-driven statistical heuristics. This might not always be the ideal approach if the use case or specific research questions can provide a more reasonable number of topics.

\subsection{Identification of biases}

\emph{RQ 2: To what extent can metadata on the persons searched (e.g., gender, age, party membership) be used to explain possible biases?}

As argued in Section \ref{sec:literature}, a bias measures a systematic deviation from an ideal or fair distribution. In the case of retrieval systems, this can be the deviation from a somehow theoretically derived fair ranking, or from an ideal retrieval set that, for example, does not contain cases of inappropriate language, hate speech or racism. The challenge here is to define and find this ideal norm. 

In our framework, we study systematic topical biases in query suggestions for groups of search terms that share similar characteristics. For instance, this can be the gender in searches for persons' names. The topics are derived from the query suggestions by using the exploratory cluster analysis. Each suggestion is labeled with a topic, and for each search term, the relative number of unique suggestions in each topical category is summed up. For specific groups of search terms, biases are then measured as significant deviations from the overall distribution of topics. Thus, the outcome of the framework is a set of statistically significant estimators that indicate a systematic deviation for particular groups. 

Whether these deviations are unfair in a theoretically justified way cannot be answered by this analysis and is subject to the researcher's interpretation. However, the proposed procedures can be applied to various search terms and are not restricted to specific use cases. They support the researcher in spotting broad biases hidden in a large amount of data. 

In our experiment, the regression analysis found some weak biases for female politicians as they receive a slightly higher proportion of query suggestion terms related to personal and emotional topics (cluster 2). Among others, another biasing factor that appeared in the data was the age. The older the politicians are, the fewer personal query suggestion terms appeared. Due to the main focus on the German Bundestag, these results are limited as we used 629 different names of politicians to form our query suggestions data set. Considering the overall weak explanatory power of our regression results, the second research question \textit{RQ 2} remains an open issue. Future research should try out more use cases and test other methods of statistical inference. For example, beta regression models are designed for the modeling of proportions and frequencies and might perform better in this case \citep{cribari2010}.

\subsection{Persistence of biases}

\emph{RQ 3: How can we measure and analyze the stability and persistence of possible biases in the suggestions?}

The bias analysis relies on a cross-section of unique search suggestions. Hence, any time dependencies are neglected here. Especially for prominent search terms, the generation of query suggestions is highly dynamic. As the retrieved lists of suggestions might change fast, future work should incorporate these dynamic effects into the analysis. For example, for each suggestion, the visibility in the ranking measured in days can be calculated. These statistics can be used as a weight when calculating the relative cluster scores such that prevalent items achieve a more significant influence.

Considering these dynamics, an important question is, whether biases in the ranking remain stable over time. In the experiment, the aggregated measures of stability showed that the rate of change in the rankings stayed relatively constant. After one year, comparing each ranking to its counterpart from the earliest time point, the rankings still showed an average similarity of roughly 45\%. There seems to be a general pattern in the data: While some trending suggestions only appeared for a short period, other suggestions were present for most of the time. It seems that these trending suggestions are often related to news events about the person searched. Studying which kind of news trigger enough searches such that the suggestion algorithms depict them might help in understanding how biases emerge online. The terms' search volume might be an essential factor in explaining the variety of long-term persistence observed. For example, when looking at the data from the experiment, it seems that well-known politicians such as Angela Merkel have a larger number of unique suggestions and more variations over time compared to other, more unknown politicians.

\section{Conclusion}
\label{sec:conclusion}
The discussion about biases is not purely academic but has got a real impact on searchers. By measuring search engine bias with retrieval measures, it was shown that fairer and less biased search systems tend to perform better \citep{wilkie_algorithmic_2017}. The design of current search engines requires users to be aware of possible biases. They must submit specific queries and thoroughly assess the result pages. As long as search engines are commercial black-box systems, this situation is unlikely to change. Throughout this article, we emphasized that research, focusing on algorithmic biases in retrieval systems, should consider the case of query suggestions. They already steer and influence the decision-making process during the early phase of query formulation. Therefore, we support the demand for a free web index \citep{lewandowski_living_2015} and wish for services and tools that simplify the study of query suggestion mechanisms. However, as a free web index is not in sight, monitoring and auditing of web search engines and their query suggestions are required to identify potential biases. 

We see our work as a first step into a systematic, algorithmic audit of query suggestions in web search engines. In a case study, we studied biases in query suggestions when searching for the names of politicians. We presented a framework, which is capable of automatically collecting and analyzing query suggestions for large sets of search terms over a long period. The suggestions are grouped into topical clusters, and possible biases can be identified. Nevertheless, it is the researchers' responsibility to interpret the topics derived from the data and the reported biases afterward.

\bibliographystyle{agsm}
\bibliography{bias2018}

\end{document}